\input{aipcheck}

\documentclass[
    ,final            
  ]
  {aipproc}

\layoutstyle{8x11double}

\def\ji{$\rm{J}_1$}
\def\jii{$\rm{J}_2$}
\def\ki{$\rm{K}_1$}
\def\kii{$\rm{K}_2$}
\def\kiv{$\rm{K}_4$}
\def\ri{$\rm{R}_1$}
\def\rii{$\rm{R}_2$}
\def\nii{[N {\sc ii}]}
\def\sii{[S {\sc ii}]}

\def\oiii{[O {\sc iii}]}

\def\cliii{[Cl {\sc iii}]}

\def\ri{$\rm{R}_1$}
\def\rii{$\rm{R}_2$}
\def\te{T$_e$}
\def\teff{T$_{eff}$}
\def\ne{N$_e$}

\begin{document}

\title{Do we really know how to derive the basic PNe parameters?}

\classification{<Replace this text with PACS numbers; choose from this list:
                \texttt{http://www.aip..org/pacs/index.html}>}
\keywords      {<Enter Keywords here>}

\author{Denise R. Gon\c calves}{
  address={IAG, Universidade de S\~ao Paulo, Rua do Mat\~ao 1226, 05508-900, 
  S\~ao Paulo, Brazil}
}

\begin{abstract}
How well do we know the physical/chemical properties of PNe? 
1D ({\sc cloudy}) and 3D ({\sc mocassin}) photoionisation codes are 
used in this contribution to model the PNe K~4-47 and 
NGC~7009 as an attempt to question whether or not the high 
Te (higher than 21,000K) of the K~4-47's core and the N overabundance 
of the outer knots in NGC~7009 are real. 

These are very basic parameters, obtained for Galactic PNe, 
e.g. nearby objects, even though with large uncertainties. 
Based on the comparison of the modelling with, mainly, optical 
images and long-slit spectroscopic data, it is suggested here that K~4-47 
high Te can be explained if its core is composed 
of a very dense and small inner region -- that matches the radio 
measurements-- and a lower density outer core --matching the optical
observations. This approach can 
account for the strong auroral emission 
lines [OIII]4363\AA\ and [NII]5755\AA\ observed, and so for the high 
temperatures. This teaches us that the assumption of a homogeneous 
distribution of the gas is completely wrong for the core of such PN. 

In the case of NGC~7009 a simple 3D model that
reproduces the observed geometry of this nebula is constructed. The aim of this modelling
was to explore the possibility that the enhanced [NII] emission observed
in the outer knots may be due to ionisation effects instead to a local 
N overabundance. Here it is discussed the model that can best reproduce the 
observations employing a {\it homogeneous} set of abundances throughout the 
nebula, not only for {\it nitrogen} but also for all the other elements considered.
\end{abstract}

\maketitle

\section{An Introduction to K~4-47 and NGC~7009}

These two PNe are similar in that both have pairs of low-ionisation 
knots and jets, the former being known as FLIERS (fast, low-ionisation 
emission regions, Balick et al. 1993). Because the fact that K~4-47 and 
NGC~7009 most appealing features are those of small-scales --knots and 
jets-- in spite of their large-scale structures --rim, shell and halo-- 
spatially resolved analysis becomes mandatory for describing them 
properly. 

A couple of years ago we (Gon\c calves et al.~2003, 2004) published the analysis 
of these PNe, 
based on their spatially resolved long-slit spectroscopic data  
--namely physical, chemical and excitation 
properties-- as part of a wider project aimed at the study of these 
properties for a large sample of PNe that are known to possess small-scale 
low-ionisation structures, LIS (see Gon\c calves~2003 for a review on 
LIS of PNe).

\begin{figure}
  \includegraphics[height=.17\textheight]{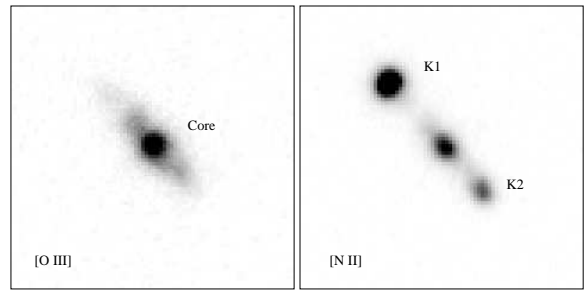}
  \caption{NOT [O {\sc iii}] and [N {\sc ii}] images of K~4-47, putting in evidence the 
prominent [N {\sc ii}] pair of knots and the [O {\sc iii}] prominent core. The
sizes of the boxes are 15$\times$15~arcsec.}
\end{figure}

As it is shown in Figure~1, K~4-47 is a compact PN, composed by a 
small, high ionisation nebular 
core and a pair of low-ionisation, high-velocity knots, connected to the core 
by a much fainter low-ionisation lane (Corradi et al.~2000).
Gon\c calves et al.~(2001) have proposed that the low-ionisation lanes
and knots of K~4-47 are genuine jets, and that their morphological and
kinematic properties can be explained if the jets and knots were
formed by accretion disks, attaining velocities of several hundred
kilometers per second. These highly supersonic velocities 
imply that the resulting LIS are likely to be shock-excited. This  
is a poorly studied PN, with statistical distances ranging from 8.5~kpc (Cahn,
Kaler \& Stanghellini 1992) to 26~kpc (van de Steene \& Zijlstra
1994).  Corradi et al. (2000) restricted this range to 3 -
7~kpc, assuming that the object participates in the ordered rotation of
the disk of the Galaxy. Tajitsu \& Tamura
(1998) estimated a distance of 5.9~kpc using the integrated {\it IRAS}
fluxes under the assumption of constant dust mass for all PNe.
Lacking of better information, we adopt the latter value in this work. 
 Lumsden et al. (2001) mapped the H$_2$ emission
from K~4-47 finding that it is excited by shocks.  The object also
appears in the 6~cm VLA radio survey of Aaquist \& Kwok (1990) showing
a very compact radio core, with a diameter of 0.25 arcsec, and one of
the largest brightness temperatures (T$_b$=8,700 K) found in PNe.  So
far, the properties (luminosity and temperature) of its central star are not
known.

\begin{figure}
 
\includegraphics[height=.18\textheight]{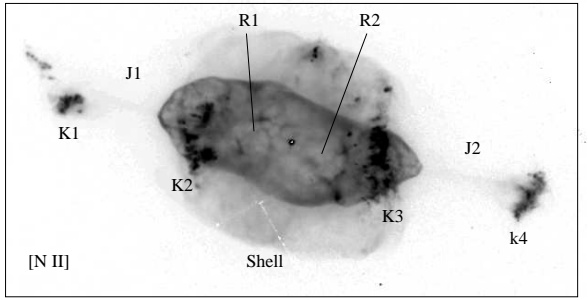}
  \caption{HST [N {\sc ii}] image of NGC~7009, in logarithmic scale. Labels 
  mark the positions of the outer (K1, K4) and inner (K2, K3) pair of knots, 
  the pair of jets (J1, J2), the rim (R1, R2) and the shell. The size of the 
  box is 65$\times$32~arcsec.}
\end{figure}

As for the ``Saturn Nebula'', NGC~7009, it comprises a bright elliptical rim 
and a tenuous halo. Its small-scale structures include a pair of jets and two 
pairs of low-ionisation knots (see Figure~2). High-excitation lines dominate the inner 
regions along the minor axis, while emission from low-ionisation species is 
enhanced at the extremities of the major axis. NGC~7009 was classified 
as an oxygen-rich PN (Hyung \& Aller~1995), with an O/C ratio exceeding 1, 
and anomalous N, O, and C abundances (Balick et al.~1994, 
Hyung \& Aller~1995). Its central star is an H-rich O-type star, with an effective 
temperature of 82,000~K (M\'endez et al.~1992).  
Reay \& Atherton (1985) 
showed that the outer knots are expanding near the plane of the sky 
at highly supersonic velocities, and that the inclination of the inner (caps) 
and outer (ansae) knots, with respect to the line of sight, are $i\cong 
51^{\circ}$ and $i\cong 84^{\circ}$, respectively.  
More recently, Fern\'andez et al.~(2004) have measured the proper motion 
and kinematics of the ansae in NGC~7009, obtaining V$_{\rm {exp}}$ = 114 
$\pm$ 32~km s$^{-1}$, for a distance of $\sim$ 0.86 $\pm$ 0.34~kpc.

\section{Empirical physical and chemical properties}

As described in Gon\c calves et al.~(2003 and 2004), the observations used 
in the present discussion include \oiii\ and \nii\ images of both PNe, 
as well as long-slit, intermediate dispersion spectra taken along the PNe 
major axes (P.A. are 41$^{\circ}$ and 79$^{\circ}$, for K~4-47 and NGC~7009, 
respectively).  

The physical parameters for these PNe were obtained from the following line 
ratios: 
\begin{enumerate}
 \item
  \sii~6717\AA/6731\AA\ and \cliii~5517\AA/5537\AA, for electron densities;
 \item
  \oiii~(4959\AA+5007\AA)/4363\AA, \nii~(6548\AA+\newline6583\AA)/5755\AA\ and 
  \sii~(6716\AA+6731\AA)/(4069\AA+\newline4076\AA), for electron temperatures.
\end{enumerate}

The total abundances for the different structures in these PNe were 
derived from an empirical analysis that uses the ionisation correction factors 
({\it icf}) to account for the unseen ions (e.g. Kingsburgh \& Barlow, 1994).
Results obtained with the {\it icf} method can be somewhat uncertain, 
particularly when they are applied to spatially resolved long-slit 
spectra (Alexander \& Balick~1997), as it is the case in the present work. 

\noindent\newline
{\bf K~4-47}: 

Electron densities and temperatures of K~4-47 are listed in Table~1. 
Both knots are denser than the core (1,900~cm$^{-3}$) by factors of 2.4 
and 1.2 for \ki\ and \kii, respectively. 
$T_e$\oiii, appropriate to zones of medium to high excitation, and $T_e$\nii\ 
characteristic of low-excitation regions are also shown, and in two cases they are 
only lower limits.  Note that temperatures of K~4-47 are remarkably higher than the
typical values for PNe (around $10^4$~K). In addition to these high
 $T_e$, diagnostic diagrams (not shown here) also suggest that shock 
excitation play an important role in the 
high-velocity knots of this PN, but not in its core. 

\begin{table}[ht]
\begin{tabular}{lrrrr}
\hline
  \tablehead{1}{l}{b}{N$_e$ (cm$^{-3}$)\\ T$_e$ (K)}
  & \tablehead{1}{r}{b}{K1\\}
  & \tablehead{1}{r}{b}{Core\\}
  & \tablehead{1}{r}{b}{K2\\}   
  & \tablehead{1}{r}{b}{Whole\\nebula}   \\
\hline
N$_e$[S{\sc ii}] &  4,600       &  1,900       & 2,400 &  2,800\\
T$_e$[O{\sc iii}]        & $\ge$21,000  & 19,300       &16,100 & 19,300\\
T$_e$[N{\sc ii}]         & 18,900       & $\ge$21,000  &16,950 & 20,600\\
\hline
\end{tabular}
\caption{Physical parameters of K~4-47}
\label{tab:a}
\end{table}

Keeping in mind that 
the Core seems to be mainly photoionised, we derive its ionic and {\it icf} 
total abundances. The derived values, with respect to H, are 
as follows: 1.39E-1, 7.37E-5, 3.74E-4, 1.74E-5 and 1.96E-6 for 
He, O, N, Ne and S, respectively. From the high He and N abundance obtained, 
K~4-47 is a typical Type~I PN of the Galactic disk, while, somewhat in contradiction, 
its extreme oxygen deficiency is typical of Galactic halo PNe.

\noindent\newline
{\bf NGC~7009}: 

\begin{table}[ht]
\begin{tabular}{lrrrrr}
\hline
    \tablehead{1}{l}{b}{N$_e$(10$^3$cm$^{-3}$)\\T$_e$(10$^3$K)}
  & \tablehead{1}{r}{b}{\ki\\}
  & \tablehead{1}{r}{b}{\ri\\}
  & \tablehead{1}{r}{b}{\rii\\}
  & \tablehead{1}{r}{b}{\kiv\\}
  & \tablehead{1}{r}{b}{Whole\\nebula} \\
\hline
N$_e$\sii 	& 2     & 5.5    & 5.9    & 1.3	 & 4\\
N$_e$\cliii 	& -	   & 5.2    & 5.9    & 1.9	 & 4.5\\
T$_e$\oiii 	& 9.6     & 10 & 10.2 & 10.4 & 10.1\\
T$_e$\nii 	& 11  & 10.4 & 12.8 & 11.7 & 10.3\\
T$_e$\sii 	& 7.1     & -       & -       & 9.4	 & -\\ \\
He/H (10$^{-1}$)& 1.0  & 1.0  & 1.1  & 0.95  & 1.1 \\ 
O/H  (10$^{-4}$)& 5.8  & 4.5  & 4.8  & 4.5  & 4.7 \\
N/H  (10$^{-4}$)& 3.8  & 0.7  & 1.8  & 2.5  & 1.7  \\   
Ne/H (10$^{-4}$)& 1.1  & 1.1  & 1.1  & 1.3  & 1.1  \\
S/H  (10$^{-6}$)& 0.13 & 6.1  & 4.9  & 9.3  & 8.3 \\
\hline
\end{tabular}
\caption{Physical parameters and abundances of NGC~7009}
\label{tab:b}
\end{table}

As one can see in Table~2, densities for the
outer knots are very similar to those of the jets connecting them
to the edge of the rim, with $N_e$(\ki) = 1,900~cm$^{-3}$
and $N_e$(\ji)= 1,300~cm$^{-3}$ and, on the opposite side,
$N_e$(\kiv) = 1,300~cm$^{-3}$ and
$N_e$(\jii) = 1,400~cm$^{-3}$. On the other hand, the  rim 
has higher electron densities, namely $N_e$(\ri)= 5,500~cm$^{-3}$ and 
$N_e$(\rii) = 5,900~cm$^{-3}$. As for the densities, the temperature 
estimators are appropriate for zones of low and medium to high 
excitation. 
The general trend of temperatures is that 
they are constant throughout the  nebula, having an average value
of $T_e$\oiii\ = 10,200~K and $T_e$\nii\ = 11,100~K. 

The empirical {\it icf} shown in Table~2 are homogeneous across the
nebula, to within 9\%, 17\%, and 35\% for He, O, and Ne and S,
respectively.  Nitrogen seems to be enhanced in the outer knots of NGC~7009 
by a factor $< 2$, but this evidence is marginal considering the 
large range in the derived {\it icf}. The uncertainties 
intrinsic to the method are also rather large (Alexander \& Balick 1997), 
but the present data seem to discard variations at the level found by Balick 
et al. (1994), namely outer knots overabundance by factors of 2 - 5. 

\section{Photoionisation Modelling}

The 1D-{\sc cloudy} (Ferland et al.~1998) and the 3D-{\sc mocassin} (Ercolano et 
al.~2003; Ercolano~2005) photoionisation codes are used here in order to check 
whether or not the high \te\ of the K~4-47's core and the N overabundance of 
the outer knots in NGC~7009 are real.

As input the photoionisation codes need information on the shape and intensity 
of the radiation from the ionizing source, the chemical composition and
geometry of the nebula, as well as its density and size.  

\noindent\newline
{\bf K~4-47}: 
As mentioned before, the distance of K~4-47, and thus the size and the
luminosity of the central star, are poorly known, but some measurements  
found in the literature lead to the values shown in the input 
parameters table. In Table~3 the L$_*$ is the lower limit 
for the star luminosity, that comes from the {\it IRAS} spectral energy
distribution (Tajitsu \& Tamura~1998). The \teff\ was obtained from the Core 
He{\sc i} and He{\sc ii} nebular emission lines. A spherical geometry is 
adopted, with the size and density (which is constant) determined from 
the optical data (Corradi et al.~2000). 
Models with the much higher density derived from the radio 
data, consistent with the more compact core, of 0.25~arcsec (Aaquist \& 
Kwok~1990) were also tested. 

With this set of parameters, we (Gon\c calves et al.~2004) found partial agreement 
with the observed 
spectrum: most of the important lines for nebular diagnosis are well reproduced,
with discrepancies of up to 35\%. However, this model underestimates, by a
factor of 3, the intensities of the auroral \oiii~4363\AA\ and \nii~5755\AA\ 
lines, that are crucial for determining \te. Note that for high density 
gas the auroral (\oiii~4363\AA\ and \nii~5755\AA) to nebular 
(\oiii~5007\AA\ and \nii~6583\AA) line ratios are indicators
of density rather than of temperature (Gurzadyan 1970). Because of that,  
models with the much higher compact core density 
(from 72,000~cm$^{-3}$ up to 10$^5$~cm$^{-3}$) are also calculated. With the latter model, the 
intensity of the auroral \oiii~4363\AA\ and 
\nii~5755\AA\ can be reproduced, despite the fact that most of the other 
lines comes to be largely underestimated because of collisional quenching.

In short, none of the constant density models is able
to account, simultaneously, for all optical emission lines in the
Core.  In particular, the \oiii~4363\AA\ and \nii~5755\AA\ intensities
are strongly underestimated if the nebular density is the one derived
empirically from the \sii\ lines.  A model with a strong density
stratification could possibly offer a solution to the problem. And, 
finally, if the strong density stratification of the Core is confirmed, 
the abundances quoted above, derived empirically, will not be valid anymore 
(Gon\c calves et al.~2004).

\begin{table}[ht]
\begin{tabular}{lrr}
\hline
    \tablehead{1}{l}{b}{Inputs}
  & \tablehead{1}{r}{b}{K~4-47}
  & \tablehead{1}{r}{b}{NGC~7009}\\
\hline
L$_*$ ($L_{\odot}$)&550      & 3,136\\
\teff\ (K)	   &120,000  & 80,000\\
\ne (cm$^{-3}$)	   &1,900    & Table~2\tablenote{We use the simplest possible
3D density distribution for NGC~7009, including an elliptical rim, surrounded by a spherical
less opaque shell. At the polar tip of the rim, we connect the cylindrical 
jets that ended as disk-shaped knots. Densities in the different volumes
(structures) match those in Table~2.}\\
D (kpc)            &5.9      & 0.86\\ 
R$_{out}$ (cm)     &4.18E+16 & 3.88E+17\\
Abundances         &Type-I PNe   & Table~2\tablenote{Except that C/H=3.2E-4;
                    N/H=2.0E-4 and Ar/H=1.2E-6}\\   
Dust grains        &ISM graphite+silicate     & -\\
\hline
\end{tabular}
\caption{Input parameters}
\label{tab:c}
\end{table}

\noindent\newline
{\bf NGC~7009}: When compared to K~4-47, the input parameters for 
the {\sc mocassin} modelling of NGC~7009 are considerably more reliable. As 
stated in the Introduction, its distance, central star effective temperature 
and luminosity are well constrained, and their values are given in Table~3. 
This table also shows the abundances adopted --mainly from Table~2 and 
Pottasch~(2000)-- and details about the geometry/density distribution assumed 
for this model. Note that: i) the emission of the spherical
shell and the elliptical rim are combined, for easy comparison with the long-slit 
spectrum;  
ii) the inner pair of knots are not being modelled, since they do not lie in the
same direction of the outer knots and will not affect very much the ionisation
structure in the outer knots; and iii) the jets are included in the model,
because they transfer the radiation from the rim to the outer knots, but as they
are very faint, their simulated spectra are not discussed.

A good agreement between the model and the dereddened intensities of the 
rim+shell and, even better, of the outer knots spectra is found (Gon\c calves et 
al.~2005). 
All model predictions fall in between the data observed at the two sides of the nebula, 
or are within 10-30\% of one of the two values. The temperature structure of 
the nebula is also well reproduced by this model, since the simulated figures 
of \te\oiii\ and \te\nii\ for the rim, the knots and the whole nebula, 
all agree with the empirical values to better than 5\%.

More importantly, concerning the ionisation structure of the simulated nebula, 
a relevant issue 
that should be emphasized here is the N/N$^+$ ratio being higher than the 
O/O$^+$ ratio by a factor of 1.39 in the knots, 1.66 in the rim and 1.61 in 
the total nebula. This result is at variance with the N/N$^+$=O/O$^+$ 
(Kingsburgh \& Barlow~1994; Perinotto et al.~2004) 
generally assumed by the 
{\it icf} method, with the consequent errors on empirically derived total 
elemental abundances. Note also that only a small fraction 
(0.7\%, 14\% and 0.6\% for rim, knots and whole nebula, respectively) of the 
total nitrogen in the nebula is in the form of N$^0$ and N$^+$. As only lines 
from these ions were observed (see Table~3 of Gon\c calves et al.~2003), the 
nitrogen abundance determination is particularly uncertain. Therefore, {\it
icf}s will be underestimated by the empirical scheme, for both components, 
rim and knots, but more so in the knots.

Finally, there is a strong dependence of the ionisation level on the geometry 
and density distribution of the gas, which makes the (N$^+$/N)/(O$^+$/O) ratio 
extremely sensitive to the shape of the local radiation field. So that a
realistic density distribution is essential to the modelling of a non-spherical 
PN, if one wants to get reliable information from spatially resolved observations 
 (Gon\c calves et al.~2005).

\section{In Conclusion}
It should be finally noted that these (electron temperatures and chemical 
abundances) diagnosis are used not only to 
the analysis of PNe, but also to symbiotic systems, HII regions, and 
other ionised nebula, for which less constraints are generally available. 
For this reason it is of vital importance to verify  the reliability of
the present techniques for the determination of the basic PNe parameters.

\begin{theacknowledgments}
DRG acknowledges the financial support of the Brazilian Agency 
FAPESP (03/09692-0, 04/11837-0) and of the Spanish Ministry of Science and 
Technology (AYA 2001-1646).
\end{theacknowledgments}

\end{document}